\documentclass[12pt,reqno]{article}

\usepackage[usenames]{color}
\usepackage{amssymb}
\usepackage{graphicx}
\usepackage{amscd}

\usepackage[colorlinks=true,
linkcolor=webgreen,
filecolor=webbrown,
citecolor=webgreen]{hyperref}

\definecolor{webgreen}{rgb}{0,.5,0}
\definecolor{webbrown}{rgb}{.6,0,0}

\usepackage{color}
\usepackage{fullpage}
\usepackage{float}

\usepackage{enumerate}
\usepackage{amsmath}
\usepackage{amsthm}
\usepackage{amsfonts}
\usepackage{latexsym}
\usepackage{epsf}
\usepackage{epsfig}
\usepackage{microtype} %

\usepackage[american]{babel}
\usepackage{braket}

\newcommand{\dff}[1]{\underset{#1}{\Delta}}
\newcommand{\E}[1]{\underset{#1}{\mathbf{E}}}
\newcommand{\M}[1]{\underset{#1}{\mathbf{M}}}

\newcommand{\1}{\mathbf{1}}

\newcommand{\oT}{\mathcal{I}}
\newcommand{\oB}{\mathcal{E}}
\newcommand{\oM}{\mathcal{M}}
\newcommand{\oD}{\mathcal{D}}
\newcommand{\oI}{\mathcal{J}}
\newcommand{\Dc}{\mathbf{D}}
\newcommand{\Dd}[1]{\underset{#1}{\mathbb{D}}}
\newcommand{\Sq}[1]{\left(#1_i\right)}
\newcommand{\iI}[1]{\sum #1\,d\mathbb{N}}
\newcommand{\dI}[3]{\sum\limits_{#1}^{#2}#3\,d\mathbb{N}}

\begin{document}

\theoremstyle{plain}
\newtheorem{theorem}{Theorem}
\newtheorem{corollary}[theorem]{Corollary}
\newtheorem{lemma}[theorem]{Lemma}
\newtheorem{proposition}[theorem]{Proposition}

\theoremstyle{definition}
\newtheorem{definition}[theorem]{Definition}
\newtheorem{example}[theorem]{Example}
\newtheorem{conjecture}[theorem]{Conjecture}

\theoremstyle{remark}
\newtheorem{remark}[theorem]{Remark}

\begin{center}
\vskip 1cm{\LARGE\bf Discrete Calculus of Finite Sequences
}
\vskip 1cm
\large
S\'ergio Martins Filho \\
Departamento de F\'isica \\
Universidade Federal de Santa Catarina \\
Florian\'opolis 88040-900 \\
Brazil  \\
\href{mailto:sergiomartinsfilho@live.com}{\tt sergiomartinsfilho@live.com} \\
\end{center}

\vskip .2 in

\abstract{The calculus of finite differences is a solid foundation for the
    development of operations such as the derivative and the integral for
    infinite sequences. Here we showed a way to extend it for finite sequences.
    We could then define convexity for finite sequences and some related
    concepts. To finalize, we propose a way to go from our extension to the
calculus of finite differences.}

\section{A little introduction on calculus of finite differences}
\label{sec:intro1}
We will introduce some definitions and notation from the calculus of finite
differences. Mostly all notation are from Jordan \cite{Jordan}, and, like the
notation, mostly definitions on the subject are the same, as we can find in
Jordan's \cite{Jordan} book or any other book on finite differences, like
Boyle \cite{Boyle}.

The definition of symbol from Boyle \cite[pp. 16--18]{Boyle} will be generalized,
and it will be used later on the study of discrete calculus of sequences.

\subsection{Difference}
The main definition of the calculus of finite differences is the difference.
The \textit{difference} of a function $f(x)$, which is given for
$x_1, x_2, \ldots, x_n$; such that $x_{i+1}-x_{i}=h$ for all $i$ between $1$
and $n-1$, is \[\dff{x, h'} f=f(x+h')-f(x).\]

\subsection{Operation of displacement}
The displacement operation is an important operation in calculus of finite
differences, and consist in increasing the argument of the function by some
amount. Then, if we denote this operation by $\E{}$, we have
\begin{equation}
  \E{h} f(x)=f(x+h).
  \label{eq:1}
\end{equation}

Note that, in $\E{}$ we are omitting $h$ and $x$ as we will do most time with
such operations, and as is vastly done in literature, like in Jordan's book
\cite[p.\ 6]{Jordan}, let us define $\E{}^n$ by
\[\E{}^nf(x)=\E{}^{n-1}[\E{} f(x)]=f(x+nh),\]
where $n$ are a positive integer, and for a negative integer $-n$, we have
\[\E{}^{-n}f(x)=f(x-nh).\]

\subsection{Operation of the mean}
The operation of the mean will be more important  later in this paper than
displacement operation, because it is invariant on the interchange of $\E{}$ and
$\1$. This operation will be denoted by $\M{}$, and is defined as follows:
\begin{equation}
  \M{h}f(x)=\frac{f(x)+f(x+h)}{2}.
  \label{eq:2}
\end{equation}

\subsection{Symbolic Calculus}
Let $\mathcal{O}$ be the set of all operation already defined with 
more one operation denoted by $\1$. Where $\1$ is the \textit{identity
operation} which take a function to itself. We note that any operation from
$\mathcal{O}$ are linear and commute with each others.

We can define addiction of operation as $[\mathbf{A}+\mathbf{B}]f(x)=
\mathbf{A}f(x)+\mathbf{B}f(x)$, for any $\mathbf{A}$ and $\mathbf{B}$, and
multiplication of a operation by a real number, in this case $\lambda$,
$[\lambda\mathbf{A}]f(x)=\lambda \mathbf{A}f(x)$, as well as multiplication of
operation that for consistency must be defined as
$\mathbf{A}\mathbf{B}f(x)=\mathbf{A}[\mathbf{B}f(x)]$.

Then we see that for operation in $\mathcal{O}$ some properties such as
associativity, distribution of multiplication over addiction, linearity and
commutativity are satisfied, moreover is easy to see that for addiction the
order of the operations does not matter.

\begin{definition}
    Let $O$ be a set of unary operations $\mathbf{O}:F\to F$. Where $F$ is any
    set of function. We say that $O$ is a \textit{symbolic set over
    $F$} when given any $\mathbf{S}$,
    $\mathbf{G}$ and $\mathbf{O}$ which belongs to $\mathcal{O}$, the following
    properties are satisfied
    \begin{enumerate}[(i)]
        \item Linearity:
        \[\mathbf{S}(\lambda f+g)=\lambda \mathbf{S}f +\mathbf{S}g,\]
        where $f$ and $g$ belong to $F$, and $\lambda$ is a real number.
    \item Commutativity (multiplication):
        \[\mathbf{S}\mathbf{O}=\mathbf{O}\mathbf{S}.\]
    \item Commutativity (addiction):
        \[\mathbf{S}+\mathbf{O}=\mathbf{O}+\mathbf{S}.\]
    \item Associativity:
        \[
        \mathbf{S}+(\mathbf{O}+\mathbf{G})=(\mathbf{S}+\mathbf{O})+\mathbf{G}.
        \]
    \item Distribution over addiction:
        \[
        \mathbf{S}(\mathbf{G}+\mathbf{O})
        =\mathbf{S}\mathbf{G}+\mathbf{S}\mathbf{O}.
        \]

\end{enumerate}
\label{def:sS}
\end{definition}

It is easy to prove that $\mathcal{O}$ is a symbolic set over the set of all
continuous functions. If $\mathbf{S}$ belongs to a symbolic set $\mathbf{O}$,
then $\mathbf{S}$ is \textit{symbol under $O$}. But, $\mathbf{S}$ still can be a
symbol under $O$ even if $\mathbf{S}$ does not belongs to $O$.

\begin{definition}
    Let $O$ be any symbolic set over $F$, and let $\mathbf{S}$ by any unitary
    operation $S:F\to F$. Hence, $\mathbf{S}$ is a \textit{symbol under $O$ over
    $F$} if the set $O'=O\cup \set{\mathbf{S}}$ is also a symbolic set over $F$.
\label{def:symbol}
\end{definition}

We already state that $\mathcal{O}$ is a symbolic set over continuous functions,
and as any differentiable function is continuous we have that $\mathcal{O}$ is
a symbolic set over differentiable functions too. If we consider the
differentiation operation denoted by $\Dc$ we can show that it is a symbol
under $\mathcal{O}$, obviously over differentiable functions.

\begin{definition}
Let $\Dd{}$ denote the \textit{discrete differentiation} operation that act in
a function as follows: \[\Dd{h} f(x)=\frac{f(x+h)-f(x)}{h}.\]
\end{definition}

First, we see that $\displaystyle \lim_{h\to 0} \Dd{h}=\Dc$, which given as a
hint about properties and interpretation of $\Dd{}$. Second, by the definition
is easy to see that $\Dd{}$ is a symbol under $\mathcal{O}$, because $\Dd{}$ is
the product of the symbol\footnote{Usually we will omit the symbolic set when
it does not cause any ambiguity.} $\dff{h}f$, and $h^{-1}$ which act as a real.
A equivalent operation is mentioned by Boyle \cite[p. 3]{Boyle} given by the
ratio $\frac{\dff{h}}{\Delta x}$, including that $\Dd{1}$ is equal to $\dff{1}$.

\begin{remark}
    It very important to remember that $\Dc$ act on any differentiable function,
    on the other hand  $\Dd{h}$ act in a wider set of function including
    differentiable function and discrete function such as sequences.
\end{remark}

\subsection{Relation between symbols under $\mathcal{O}$}
Now is clear that with respect to addiction, subtraction and multiplication, as
we defined for any operation, the symbols works as algebraic quantities. Then
we can find relation between symbols, and obtain more symbols by addiction and
multiplication of symbols.

We can give some well know equalities as follows:
\begin{align}
  \dff{}&=\E{}-\1,
  \label{eq:3}\\
  \M{}&=\frac{\1+\E{}}{2}\quad\text{and}
  \label{eq:4}\\
  \Dd{h}&=h^{-1}\dff{h}.
  \label{eq:5}
\end{align}

\section{Finite Sequences}
\label{sec:fs}

Now we will work mostly with finite integer sequences. For clarity let us define
addiction and multiplication of sequences:
\[(S+G)(i)=S(i) +G(i)\text{,}\]
\[(SG)(i)=S(i)G(i),\]
we also can define multiplication by a real number; $(\lambda S)(i)=
\lambda S(i)$, where $\lambda \in \mathbb{R}$.


\subsection{Top, middle and bottom operations}
We let $\oT$ denote \textit{top operation} that is defined by cutting off the
last term of a $n$-tuple as result we obtain a subsequence of length $n-1$.

\begin{definition}
  Let $S=\Sq{s}$ be a finite sequence of length $n>0$. The \textit{top operation}
  is defined as follows:
  \[[\oT S](i)=s_i\quad\text{ for all }i\in\set{1,\ldots,n-1}.\]
  The resulting subsequence $\oT S$ is called the \textit{top} of $S$.
\end{definition}

Meanwhile, the \textit{bottom operation} $\oB$ is defined by cutting off the
first term of a $n$-tuple resulting in a subsequence of length $n-1$.

\begin{definition}
  Let $S=\Sq{s}$ be a finite sequence of length $n>0$. The
  \textit{bottom operation} is defined as follows
  \[[\oB S](i)=s_{i+1}\quad\text{for all }i\in\set{1,\ldots,n-1}. \]
  The resulting subsequence $\oB S$ is called the \textit{bottom} of $S$.
\end{definition}
\begin{remark}
 The bottom operation shift each term to the following term, and act like $\E{1}$
for finite sequences.
\end{remark}

The \textit{middle operation} denoted by $\oM$ act in a sequence $S$ resulting
in the mean of the top and the bottom of $S$.

\begin{definition}
  The \textit{middle operation} $\oM$ is defined by his action in a sequence
  $S=\Sq{s}$ of length $n>0$ as follows:\[[\oM S](i)=\frac{s_i+
  s_{i+1}}{2}\quad\text{for all }i\in\set{1,\ldots,n-1}.\]
  The subsequence $\oM S$ is called the \textit{middle} of $S$.
  \label{def:mO}
\end{definition}

\begin{remark}
  If a sequence $R$ have a unit length. Hence $\oT R$, $\oB R$ and $\oM R$ are
  equal to the empty sequence that we will denote by $(\varnothing)$.
  None of the above operation are defined for empty sequences, but its definitions can be extend, such that,
  \[\oT (\varnothing)=\oM (\varnothing) =\oB (\varnothing)=(\varnothing).\]
\end{remark}

We already point out that $\oM S =\frac{1}{2}(\oB S +\oT S)$. If one compare it with
\eqref{eq:4} is clear that $\oM$ act in a finite sequence as $\M{}$.
At the same time is clear that $\1$ does not act as $\oT$, clearly $\1 S =S$
meanwhile the top of $S$ have a different length. Then we need the top operation
$\oT$ to work with finite sequences.

\subsubsection{Properties}
For all similarity with symbols under $\mathcal{O}$, such as $\E{}$ and $\M{}$,
if we let $O=\set{\1, \oT, \oB, \oM}$. Then, for any finite sequence $S$ and
$G$ of same length, we have some properties as follows, where $\mathbf{P}$,
$\mathbf{Q}$ and $\mathbf{R}$ belongs to $O$:

\begin{enumerate}[(i)]
  \item Linearity: For any $\lambda$ a real number, we
    have
    \[\mathbf{P}(\lambda S+G)=\lambda \mathbf{P}S +\mathbf{P}G.\]
      \begin{proof}
        We just need to show that $\oT$ and $\oB$ are linear. By definition,
        we have
        \[
        \oT\left[\lambda S+G\right](i)=\lambda S(i)+G(i)=\lambda\, \oT S+\oT G
        \quad\text{for $i = 1, \ldots, n-1$;}
        \]
         and for $\oB$ is almost the same:
       \[
       \oB[\lambda S+G](i)=\lambda S(i+1)+G(i+1)=\lambda\,\oB S+\oB G
       \quad\text{for $i = 1, \ldots, n-1$.}\qedhere
       \]
      \end{proof}
    \item Commutativity (multiplication):
      \[\mathbf{P}\mathbf{Q}=\mathbf{Q}\mathbf{P}.\]
      \begin{proof}
      Proving that $\oT$ and $\oB$ commute are enough as $\oM$ is a linear
      combination between them, and $\1$ is know to commute with any operation.
      So, by multiplication definition
      \[\oT\oB S(i) = \oT[S(i+1)]=S(i+1)\quad
    \text{where $i = 1,\dots,n-2$, and}\]
      \[\oB\oT S(i) = \oB[S(i)]=S(i+1)\quad\text{where $i = 1,\dots,n-2$.}\]

      Therefore, $[\oT, \oB] S=0$.
      \end{proof}
    \item Commutativity (addiction):
      \[
      \mathbf{P}+\mathbf{Q}=\mathbf{Q}+\mathbf{P}.
      \]
      \begin{proof}
      For real sequences and by definition of addiction of sequences is clear
      that the commutativity for addiction holds.
    \end{proof}
        \[
        \mathbf{P}+(\mathbf{Q}+\mathbf{R})=(\mathbf{P}+\mathbf{Q})
        +\mathbf{R}.
        \]
    \begin{proof}
      It is also trivial to prove that associativity is valid for real sequences.
    \end{proof}
      \[
      \mathbf{P}(\mathbf{Q}+\mathbf{R})=\mathbf{P}\mathbf{Q}
      +\mathbf{P}\mathbf{R}.
      \]
      \begin{proof}
      We have that, for example,
      \[
      \oT(\oB+\oM)S(i)=\oT\left[S(i+1)+\frac{S(i)+S(i+1)}{2}\right]=\oT\oB
      +\oT\oM\quad\text{for $i=1,\ldots,n-2$.}
      \]
      It is similar for others combination of symbols.
      \end{proof}
\end{enumerate}

We see by Definition~\ref{def:sS} that the set $O$ is a symbolic set over the
set of all finite sequences denoted by $\mathcal{F}$. In the same way we studied in
Section~\ref{sec:intro1} that symbols works as algebraic quantities, and we
could have relation between them. We have, for example, by
definition~\ref{def:mO} that
\begin{equation}
  \oM=\frac{\oT+\oB}{2}.
  \label{eq:8}
\end{equation}

\section{Derivative of finite sequences}
\label{sec:DofS}
A good way to define the derivative would use the know discrete derivative. Let
$S$ be any finite sequence of length equal to $n$. If $\Dd{}$
is applied to $S$ this would result in a sequence of length equal to $n-1$,
and $\Dd{}S(i)$ would be equal to $S(i+1)-S(i)$ for $0<i<n$.

\begin{definition}
  The \textit{derivative of a sequence} $S$ denoted by $\oD S$ is defined as
  follows:
  \begin{equation}
    (\oD S)(i)=\Dd{1} S(i)=S(i+1)-S(i)\quad\text{$i=1, 2, \ldots, (n-1)$;}
    \label{eq:6}
  \end{equation}
  where $n$ is the length of $S$. The derivative is the result of the
  differentiation operation denoted by $\oD$.
  \label{def:dS}
\end{definition}

We will show later that $\oD$ is very similar to the usual differentiation. We
will find product and quotient rules like in calculus, and use first and second
derivatives just like in calculus to classify sequences.

With the use of bottom and top operation we have that differentiation can be
defined as follows
\begin{equation}
  \oD =\oB -\oT.
  \label{eq:7}
\end{equation}
\begin{proof}
  \[(\oB - \oT)S(i)=(\oB S)(i)-(\oT S)(i)=S(i+1)-S(i)=(\oD S)(i) \qedhere\]
\end{proof}
The equality \eqref{eq:7} is the equivalent of the equality \eqref{eq:3} for
finite sequences. Still by \eqref{eq:7} we have that $\oD$ is a symbol under
$O$, because $\oD$ is a linear combination of the symbols $\oT$ and $\oB$.

\subsection{Differentiation rules}
Beside all general properties that $\oD$ already satisfy as a symbol under $O$.
We can obtain some important equalities that make clear the relation between
differential calculus, calculus of finite differences and discrete calculus
of sequences.

\begin{lemma}
  Let $S$ be a sequence. The derivative of $S$ is equal to a constant sequence
  $\oD S =(0_i)$, if and only if, $S$ is a constant sequence.
  \label{lemma:1}
\end{lemma}
\begin{proof}
  If $S$ is a constant sequence, then $S(i+1)=S(i)$. Therefore, $(\oD S)(i)=
  \Sq{0}$. Meanwhile if $(\oD S)(i)=0$, then $S(i+1)-S(i)=0$. Therefore, $S$ is
  a constant sequence as we want to prove.
\end{proof}
Below we listed some remarks regarding the derivative of sequences:
\begin{enumerate}[(i)]
  \item Let $\lambda \in \mathbb{R}$, hence $\oD \Sq{\lambda} = \Sq{0}$. We
    already proved it in the first part of the proof of the Lemma~\ref{lemma:1}.
  \item If $S$ is a arithmetics sequence with common difference $d$. Hence
  $\oD S=\Sq{d}$. In the other way, if $\oD S =\Sq{d}$ and $d$ is a real number
    different from $0$. Then, $S$ is a arithmetics progression with common
    difference equal to $d$. We will omit the proof, because it is obvious from
    Definition~\ref{def:dS} and the definition of arithmetics progression.

  \item A geometric sequence $S$ with common ratio $q\neq0$ is a solution of the following equation $\oD S =(q-1)\oT S$.
    \begin{proof}
      We have, by definition of geometric progression, that
      \[
      \frac{\oB S}{\oT S}=q \Rightarrow \frac{\oB S-\oT S}{\oT S}=
      \left( [q-1]_i \right),
      \]
      therefore $\oD S = (q-1)\oT S$, remembering that the product of a constant
      sequence $\Sq{\lambda}$ and any sequence of same length $S$ is equal to
      the product of $S$ and the real number $\lambda$.
    \end{proof}
    Let us take a real function $f$, such as, $f(j)=a_1q^j$, where $j$, $a_1$
    and $q$ are real numbers. We have that the derivative is
    \[f'(j)=f(j)\log{q},\]
    and the discrete derivative is
    \[\Dd{1}f(j)=f(j+1)-f(j)=a_j(q-1).\]

    Hence, we note that studying the discrete derivative of a real function given
    us a way to find the derivative for the sequence of images $f(i)$ with
    $i=i_0, i_0+1\cdots$. 
  \item Let $S$ be a sequence, such as, $S(i)\neq0$ for all $i$. We let $S^{-1}$
    denote the inverse of $S$ which is defined as follows:
    \[S^{-1}(i)=\frac{1}{S(i)}.\] So, the derivative of $S^{-1}$ is
    \[
    \oD S^{-1}(i)=\frac{1}{S(i+1)}-\frac{1}{S(i)}=\frac{S(i)-S(i+1)}{S(i)S(i+1)}.
    \]

    Finally, we have that
    \[\oD S^{-1}=-\frac{\oD S}{\oT S\,\oB S},\]
    which looks like the derivative of the inverse of a function $f$ given by
    $-\frac{f'}{f^2}$.
\end{enumerate}

\subsubsection{Higher order derivatives}

From \eqref{eq:7} we must have that
\begin{equation}
\oD^m=(\oB-\oT)^m=\sum_{k=0}^{m}(-1)^{k}\binom{m}{k}\oT^k\oB^{m-k}.
  \label{eq:hoD}
\end{equation}

Hence, for example,
\begin{equation}
  \oD^2 S(i)=1S(i+2)-2S(i+1)+1S(i),
\label{eq:2D}
\end{equation}
with \eqref{eq:8} is possible obtain others equations for higher order of $\oD$
that can be useful in some situations.

\subsubsection{Product rule}
Let $S$ and $G$ be two sequences of same length. The derivative of $SG$ is,
by \eqref{eq:7}
\begin{align}
  \oD(SG)=\oB(SG)-\oT(SG)&=\oB{S}\,\oB{G}-\oT{S}\,\oT{G},
  \label{eq:10}\\
               \oD(SG)  &=\oD{S}\,\oB{G}+\oT{S}\,\oD{G}\quad\text{and}
  \label{eq:11}\\
               \oD(SG)  &=\oD{S}\,\oT{G}+\oB{S}\,\oD{G}.
  \label{eq:12}
\end{align}

The equality \eqref{eq:10} can be easy proved as the generalization
\[\mathbf{O} \prod^{n}_{i=1}S_i=\prod^{n}_ {i=1}\mathbf{O} S_i,\] where
$\mathbf{O}$ can be both $\oT$ or $\oB$. Then, we obtain that $\oD(SG)$ is equal
to \eqref{eq:11} and \eqref{eq:12} which are similar to product rule from
usual calculus. However, \eqref{eq:11} and \eqref{eq:12} are not symmetric.

It is not hard to see that if we change $\oT$ by $\oB$ in \eqref{eq:11} as
result we would have \eqref{eq:12}, and vice versa. Now, let us sum the
equations \eqref{eq:11} and \eqref{eq:12}, and rewrite $\oD(SG)$ as follows:
\begin{equation}
  \oD(SG)=\oD S\,\oM G + \oM S\, \oD G.
  \label{eq:prodR}
\end{equation}

Equation \eqref{eq:prodR}, differently of \eqref{eq:11} and \eqref{eq:12}, is
symmetric, and analogous to the usual product rule from differential calculus.
Moreover, the \emph{product rule} \eqref{eq:prodR} is equal to the product
rule of calculus of finite differences, where we have
\[\dff{}(fg)=\dff{}f \M{}g+\M{}f\dff{}g.\]

\subsubsection{Quotient rule}
It is easy to show that if $S$ and $G$ are sequences of same length, then
\[
\oD \left( \frac{S}{G} \right)=\frac{\oD S\,\oM G-\oD G\,\oM S}{\oT G\,\oB G}.
\]
We have two way to show it: go for it straight by Definition~\ref{def:dS} or use
the product rule that we obtained above.

Using the equality proved in the item (iv) at the beginning of the
section, \[\oD G^{-1} =-\frac{\oD G}{\oT G\,\oB G},\] and with the product
rule, we have
\[
\oD(SG^{-1})=\oD S\, \oM G^{-1}+\oM S \, \oD{G^{-1}}=
\frac{\oD S \,\oM G^{-1}\,\oT G\,\oB G-\oM S\,\oD G}{\oT G\,\oB G}.
\]

Finally, by definition of the middle operation:
\[
\left( \oM G^{-1} \right)(i)=\frac{1}{2}\left(
\frac{1}{G(i+1)}+\frac{1}{G(i)} \right)=\frac{(\oM G)(i)}{(\oT G)(i)(\oB G)(i)},
\]
thus we have the quotient rule as expected
\begin{equation}
  \oD \left( \frac{S}{G} \right)=\frac{\oD S\,\oM G-\oD G\,\oM S}{\oT G\,\oB G}.
  \label{eq:quoR}
\end{equation}

\section{Integral of a finite sequence}
\label{sec:IofS}
Now with a definition for derivative of sequences we can try now look
for a inverse of the  differentiation operation. We expect, if it exists, be
some operation like integration in calculus.

\subsection{Indefinite integral and integration operation}
Let $\oI$ denote the operation of \textit{integration} of sequences defined as
the inverse of the derivative $\oD$. Hence, we have
\begin{equation}
  \oI\oD=\oD\oI=\1.
  \label{eq:defI}
\end{equation}

If we take the derivative of a $n$-tuple $S$, and then integrate we obtain the
following equality
\[(\oI\oD S)(i)=\oI(S(i+1)-S(i))=S(i)=\1S.\]

Therefore we have that
\[(\oI S)(i+1)=(\oI S)(i)+S(i),\]
finally we can conclude that
\[
(\oI S)(i)=(\oI S)(1)+\sum\limits^{i-1}_{j=1}S(j)\quad\text{for $1<i\le n$.}
\]

The value $(\oI S)(1)$ is equivalent to the addiction of a constant sequence to
$\oI S$ like in the indefinite integral in differential calculus.

\begin{definition}
  We let $\iI{S}$ denote the \textit{indefinite derivative} of the $n$-tuple
  $S$ that is defined as the following sequence:
  \begin{equation}
    \left(\iI{S}\right)(i)=\left(\iI{S}\right)(1)+\sum\limits^{i-1}_{j=1}S(j)
    \quad\text{for $1<i\le n+1$.}
    \label{eq:9}
  \end{equation}
  Where $\left(\iI{S}\right)(1)$ is some real number.
  \label{def:1}
\end{definition}

\begin{theorem}
  The indefinite integral of the sequence $S$ is equal to the resulting sequence
  of the integration of $S$.
  \label{theo:1}
\end{theorem}
\begin{proof}
  By definition $\oI(\oD S)=\1$. Hence, we only need to prove that $\oD$ and
  $\oI$ commute. The derivative of $\iI{S}$, where $S$ is a sequence of
  length $n$, from \eqref{eq:9}, is
  \[
  \left(\oD \iI{S}\right)(i) = \sum\limits^{i}_{j=i}S(j)=S(i)=(\1 S)(i).\qedhere
  \]
\end{proof}
\begin{remark}
  The bottom of the indefinite integral of a sequence $S$ is equal to the
  sequence of partial sum of the elements of $S$ up to a constant sequence.
  \label{rm:1}
\end{remark}
\subsection{Integration Rules}
It is not hard to prove that $\oI$ is a symbol under $O$, and, like $\oD$, the
integration of sequences looks like the integration of Calculus.

\begin{lemma}
  Let $S$ be a sequence. The indefinite integral of $S$ is a constant sequence,
  if and only if, $S$ is equal to $\Sq{0}$.
  \label{lemma:2}
\end{lemma}
\begin{proof}
  It is easy to see, see Remark~\ref{rm:1}, that $\iI{\Sq{0}}=\Sq{\lambda}$ where
  $\lambda$ is some real number. Meanwhile if $\iI{S}=\Sq{\lambda}$ then by
  Theorem~\ref{theo:1}, we know that
  \[\oD\iI{S}=S=\oD\Sq{\lambda},\]
  therefore by Lemma~\ref{lemma:1} follows that  $S=\Sq{0}$.
\end{proof}
\subsubsection{Integration by parts}
Now that we have a product rule let us integrate \eqref{eq:prodR}, like we would
do in calculus to find the rule of integration by parts, we have by the
definition of $\oI$:
\[
  SG=\oI\left(\oD S\,\oM G\right)+\oI\left(\oM S\,\oD G\right),
\]
using the Theorem~\ref{theo:1} we can rewrite the above equation as follows:
\begin{equation}
  \iI{\oD S\,\oM G}=SG-\iI{\oM S\,\oD G}.
  \label{eq:13}
\end{equation}

Above equation works as the integration by parts of calculus. Moreover, if
$S$ is a sequence and we let $dS\equiv\oD Sd\mathbb{N}$, then \eqref{eq:13} is
even more similar:\[\sum\oM S\,dG=SG-\sum \oM G\,dS.\]

\subsection{Definite integral}
As we have in calculus, we will define a real number called definite integral.
A theorem like the second fundamental theorem of calculus should be satisfied.

\begin{definition}
  Let $S$ be a sequence of length $n$. If $a$ and $b$ are two integers between
  $0$ and $n+1$, then the \textit{definite integral} from $a$ to $b$ of $S$ is
  denoted by $\dI{a}{b}{S}$. Where \[\dI{a}{b}{S}=\sum\limits^{b}_{j=a}S(j).\]
\end{definition}

The definition given above is intuitive taking in consideration that a sum is
the discrete version of the integral\footnote{As a Riemann Integral.}. We see that the definite integral of the
derivative of a $n$-tuple $S$ from $1$ to $n-1$ is
\[\dI{1}{n-1}{\oD S}=\sum\limits^{n-1}_{j=1}(S(j+1)-S(j))=S(n)-S(1),\]
then we can state a theorem like the second fundamental theorem of the calculus.

\begin{theorem}[Second Fundamental Theorem of Discrete Calculus of Sequences]
  Let $S$ be a sequence with $n$ terms, and let $I$ be a sequence which derivative
  $\oD I$ is equal to $S$. If $a$ and $b$ are integers greater than $0$ and
  less than or equal to $n$, then
  \[\dI{a}{b}{S}=I(b+1)-I(a).\]
  \label{theo:2}
\end{theorem}
\begin{proof}
  If $\oD I = S$, then $I=\oD^{-1}S$. Hence, from \eqref{eq:defI} we have that
  $I$ is the result of the integration of $S$. Thus, we only have to show that
  $\oI S\rvert^{b+1}_{a}\equiv \oI S(b+1)-\oI S(a)$ is equal to the definite
  integral of $S$ from $a$ to $b$.

  By Theorem~\ref{theo:1} and Definition~\ref{def:1}, we have
  \[
  \oI S\vert^{b+1}_{a}=\sum\limits^{b}_{j=1}S(j)-\sum\limits^{a-1}_{j=1}S(j)
  =\sum\limits^{b}_{j=a}S(j)=\dI{a}{b}{S}.
  \]
  Since $I=\oI S$,
  \[\dI{a}{b}{S}=I\rvert^{b+1}_{a}=I(b+1)-I(a).\qedhere\]
\end{proof}

Some results are listed below.
\begin{enumerate}[(i)]
  \item Let $S$ be a geometric progression of length $n$. The sequence $S$
    satisfy the discrete equation $\oD S=(q-1)\oT S$, where $q$ is the common
    ratio of $S$. Taking the integral of the discrete equation from $1$ to $n-1$,
    we have
    \[
      \dI{1}{n-1}{\oT{S}}=\frac{1}{q-1}\sum\limits^{n-1}_{j=1}S(j)=\frac{S(n)-S(1)}{q-1}
      =S(1)\frac{1-q^{n-1}}{1-q}.
    \]
  \item If $S$ is a arithmetics progression of length $n$ and common difference
    $d$. Then $\oD S =\Sq{d}$, let integrate it from $1$ to $i\le n$ as follows:
    \[S(i+1)-S(1)=(i-1)d\therefore S(i+1)=S(i)+(i-1)d.\]
\end{enumerate}
\section{Application of the derivative of sequences}

\subsection{Increasing and decreasing sequences}
The Lemma~\ref{lemma:1} tell
us that in case the first derivative of $S$ is $\Sq{0}$ then $S$ does not
increase nor decrease. Hence to cover all possibilities we have that:
\begin{enumerate}[(i)]
  \item If $\oD S(i)>0$ for all $i$, then $S$ is \textit{strictly monotonically
    increasing}.
  \item If $\oD S(i)<0$ for all $i$, then $S$ is \textit{strictly
      monotonically decreasing}.
  \item If $\oD S(i)\ge0$ for all $i$, then $S$ is \textit{monotonically
    increasing}.
  \item If $\oD S(i)\le0$ for all $i$, then $S$ is \textit{monotonically
    decreasing}.
\end{enumerate}

Exactly as we do in calculus for functions using the first derivative.

\subsection{Convexity}
The calculus of finite differences is a well know way to generalize the
conception of convexity of sequences. It is done in many papers
like~\cite{MandM} and~\cite{Toader}. We will use this concept with respect to
the second derivative that can be take as the second difference $\dff{}^2$.

\begin{definition}
  A sequence $S$ is \textit{convex} when the second derivative of $S$ is
  monotonically increasing. However, if $- S$ is convex,
  then $S$ is \textit{concave}.
  \label{def:covex}
\end{definition}

We see that $\oD^2$ given us information about a sequence like the second
derivative $\Dc^2$ in calculus. The geometric interpretation of convexity is
clear let take the term of the second derivative of a sequence $S$ which is
given by \eqref{eq:2D}:
\[\oD^2S(i)=S(i+2)-2S(i+1)+S(i)\quad\text{for all i,}\]
if $S$ is, for example, convex. We have that
\[S(i+1)\le\frac{S(i+2)+S(i)}{2}\quad\text{for all i}\le n-3,\]
where $n$ is the length of $S$, then the central term $S(i+1)$ is less than the
mean of the extreme terms.

In the other hand, let take the graph of the sequences which is given by the
pairs $(j,S(j))$, a point in a Cartesian coordinate system. We note that the
point of the central term $\left( i+1, S(i+1) \right)$ is always below to the
line passing through extremities points $(i,S(i))$ and $(i+2,S(i+2))$, as we
have for continuous functions.

Moreover, let $L_2(x)=ax^2+bx+c$ be a quadratic function such that $L_2(i)=S(i)$,
$L_2(i+1)=S(i+1)$ and $L_2(i+2)=(i+2)$.The polynomial $L_2$ exists, for all $i$
 greater than 0 and less than or equal to $n-2$,
only if $\oD S(i)\neq0$, and $\oD^2 S(i) \neq0$. It is easy to check
that, if exists, $L_2(x)$ is given by
\[
    L_2(x)=\frac{(x-i-1)(x-i-2)}{2}S(i)-\frac{(x-i)(x-i-2)}{1}S(i+1)
    +\frac{(x-i)(x-i-1)}{2}S(i+2).
\]

The second derivative is
\[
  L_2''(x)=S(i+1)+S(i)-2S(i+1)=2a=\oD^2 S(i),
\]
so $L_2$ is convex (concave), if and only if, $S$ is convex (concave).

The function $L_2$ is a good way to understand the convexity of sequences,
and we will study it more later. Now going back
to a more geometric view of convexity. Let us calculate the area of the triangle
whose vertices are $(i+j,S(i+j))$, where $i$ is a integer greater than $0$ and
less than to the length of $S$ minus one, and $j= 0,1,2$. Let $A$ be the following
determinant
\[
  A=
  \begin{vmatrix}
    i & S(i) & 1 \\
    i+1 & S(i+1)  &1 \\
    i+2 & S(i+2) &1
  \end{vmatrix}=S(i)+S(i+2)-2S(i)=\oD^2 S(i),
\] we know that $\frac{1}{2}\rvert{A}\rvert$ is equal to the area of the triangle,
so the area of the triangle is proportional to the absolute value of the
second derivative. The determinant $A$ defined in a natural way can be viewed as
the value of a oriented area, when positive $S$ is convex, and when negative $S$
is concave. Beside, if $A'$ is the determinant $A$ for $-S$, we have that $A'=-A$,
in agreement with the definition.

\begin{definition}
  A sequence $S$ is \textit{strictly convex} if the second derivative of $S$ is
  strictly monotonically increasing. But, if $-S$ is strictly convex, hence
$S$ is \textit{strictly concave}.
\end{definition}

All statements below are equivalent:

\begin{enumerate}[(i)]
  \item The sequence $S$, with $n$ terms, is strictly convex (concave).
  \item The three points $(i,S(i))$, $(i+1,S(i+1))$ and $(i+2,S(i+2))$ are not
    collinear, for all $i$ greater than $0$ and less than $n-1$.
  \item The determinant $A$ is greater (less) than zero, for all $i$ such
    that $0<i\le n-2$.
\end{enumerate}
\begin{proof}
  It is well know that (iii) $\Rightarrow$ (ii). If (ii) is true, we have that
  $S(i+1)\neq\frac{1}{2}(S(i)+S(i+2))$ for all $i$, hence $\oD^2 S(i)\neq 0$ for
  all $i$. Thus (ii) $\Rightarrow$ (i). Finally, we note that if (iii) is false
  then $\det A =0$ for some $i$. So there is one $i$ such that $\oD^2 S(i) =0$.
  Therefore (i) $\Rightarrow$ (iii).
\end{proof}
\begin{definition}
  Let $S$ be a sequence of length greater than or equal to $3$. Hence $S$ is
  \textit{continuously convex}, if and only if, $S$ is strictly convex and $\oD
  S(i)\neq 0$ for all $i$ greater than $0$ and less than the length of $S$.
  If $-S$ is continuously convex, then $S$ is \textit{continuously concave}.
\end{definition}

\section{Lagrange polynomials}

Let $S_n:I_n\to\mathbb{R}$, where $I_n=\set{x\in\mathbb{N}:x\le n}$. Consider
$m+1\le n$ consecutive points $(i,S_n(i))$ with $i = n_0,\ldots, n_0+m$ where
$1\le n_0\le n-m$. The Lagrange polynomial is given by
\begin{equation}
  L(x;n_0,m)=\sum_{j=n_0}^{n_0+m}S_n(j)\prod_{\substack{k=n_0 \\ k\neq j}}^{n_0+m} \frac{x-k}{j-k}.
  \label{eq:Lpoly}
\end{equation}

We used $L(x;i,2)$ to study convexity, the function $L_2$. The reason for it came
from the general fact that
\[ \frac{d^m}{dx^m}L(x;n_0,m)=\oD^m S_n(n_0).\]

\begin{proof}
  From Eq. \eqref{eq:Lpoly}, we have
  \[
    \frac{d^m}{dx^m}L(x;n_0,m)=\sum_{j=n_0}^{n_0+m}S_n(j) \frac{d^m}{dx^m}
    \prod_{\substack{k=n_0 \\ k\neq j}}^{n_0+m} \frac{x-k}{j-k},
  \]
  now using the general Leibniz rule for the derivative of the product:
  \[
   f_m=\frac{d^m}{dx^m} \prod_{\substack{k=n_0 \\ k\neq j}}^{n_0+m}
   \frac{x-k}{j-k}=\sum_{w_1+w_2+\cdots+w_m=m}^{}
   \frac{m!}{w_1!w_2!\cdots w_m!}\prod_{\substack{k=n_0 \\ k\neq j}}^{n_0+m}
   \frac{d^{w_k}}{dx^{w_k}}\frac{x-k}{j-k}
  \]

  From the product of derivatives of linear functions we see that $w_i$ is
  equal to
  $0$ or $1$, for all $i= 1, \ldots, m$. Then with the constraint that the sum
  of all $w$'s must be $m$ follows that $w_i=1$ for all $i$. So now $f_m$ can
  be simply evaluated as
  \[
   f_m= m!\prod_{\substack{k=n_0 \\ k\neq j}}^{n_0+m} \frac{1}{j-k}=(-1)^{m+j-n_0}{m \choose j-n_0}.
  \]

  Thus
  \[
    \frac{d^m}{dx^m}L(x;n_0,m)=\sum_{j=n_0}^{n_0+m}(-1)^{m+j-n_0}
    {m\choose j-n_0}S_n(j),
  \]
  it is clear from \eqref{eq:hoD} that
  \[ \oD^mS(n_0)=(-1)^m\sum_{k=0}^{m}(-1)^{k}\binom{m}{k}\oT^{m-k}\oB^{k}S_n(n_0)
      =\sum_{j=n_0}^{m+n_0}(-1)^{m+j-n_0}\binom{m}{j-n_0}S_n(j),
  \]
  therefore, we have
\[ \frac{d^m}{dx^m}L(x;n_0,m)=\oD^m S_n(n_0).\]
\end{proof}
The order of $L_m\equiv L(n_0,m)$ is less than or equal to $m$. Then, we have that
\[L(x;n_0,m)=\sum_{i=0}^{m}l_ix^i,\]
where $l_i$ depends on $i$ and the terms of $S$. The order of $L_m$ can
be find as the maximum integer j, such that, $l_j \neq 0$.

From definition, we must have the following equality
\[ \sum_{i=0}^{m}l_1x^i=\sum_{j=n_0}^{n_0+m}S_n(j)\prod_{\substack{k=n_0 \\ k\neq j}}^{n_0+m} \frac{x-k}{j-k}, \]
after differentiating $m$ times, we obtain \[ l_m=\frac{1}{m!}\oD^mS_n(n_0).\]

Thus the order of the polynomial $L(n_0,m)$ is $m$, where $m$ is the index that
identify $L$, if and only if, $\oD^m S(n_0)\neq 0$. This explain, why $L_2$
exists as a quadratic function, if and only if, the sequence associated with $L$
is strictly convex (concave).

We also can generalize the result of the determinant $A$ for higher order of
derivative. This fact comes from the polynomial interpolation of a set of data.
For a set as the graph of a sequence $S_n$, where $n>m$, the following system
of equation must be satisfied
\[\begin{bmatrix}
i^m  & i^{m-1} & i^{m-2} & \ldots & i & 1 \\
(i+1)^m  & (i+1)^{m-1} & (i+1)^{m-2} & \ldots & i+1 & 1 \\
\vdots & \vdots    & \vdots    &        & \vdots & \vdots \\
(i+m)^m  & (i+m)^{m-1} & (i+m)^{m-2} & \ldots & i+m & 1
\end{bmatrix}\begin{bmatrix} l_m \\ l_{m-1} \\ \vdots \\ l_0 \end{bmatrix}  =
\begin{bmatrix} S(i) \\ S(i+1) \\ \vdots \\ S(i+m) \end{bmatrix},\]
where $i$ is any integer less than or equal to $n-m$.

The solution for $l_m$ is given by Cramer's rule as follows
\[ l_m=\frac{1}{m!}\begin{vmatrix}
S(i)^m  & i^{m-1} & i^{m-2} & \ldots & i & 1 \\
S(i+1)^m  & (i+1)^{m-1} & (i+1)^{m-2} & \ldots & i+1 & 1 \\
\vdots & \vdots    & \vdots    &        & \vdots & \vdots \\
S(i+m)^m  & (i+m)^{m-1} & (i+m)^{m-2} & \ldots & i+m & 1
\end{vmatrix},
\]
therefore we have
\[ \oD^m S_n(i)=\begin{vmatrix}
S(i)^m  & i^{m-1} & i^{m-2} & \ldots & i & 1 \\
S(i+1)^m  & (i+1)^{m-1} & (i+1)^{m-2} & \ldots & i+1 & 1 \\
\vdots & \vdots    & \vdots    &        & \vdots & \vdots \\
S(i+m)^m  & (i+m)^{m-1} & (i+m)^{m-2} & \ldots & i+m & 1
\end{vmatrix}.
\]

\section{Discrete calculus for infinite sequences}
In the previous sections we studied how to find a discrete derivative for
finite sequences, and we saw that the calculus of finite differences gives a
good foundation for it. For infinite sequences we can simply apply any concept
of the calculus of finite differences. We will try make it clear as follows
extending the top operation for infinite sequences.

\subsection{Top operation action on finite sequences}
It's clear that if $S$ is a infinite sequence, then the top of $S$ is simply
itself, that is,
\[ \oT{S}=\1 S.\]

Let $S$ be infinite sequence, and let $S_n$ be the sequence of the first $n$
terms of $S$. Note that $\oT{S_n}=S_{n-1}$. We can state that
\[\lim_{m \to \infty} S_m = S,\]
in this sense, we can say that:
\[\lim_{m \to \infty} \oT{S_m} =\lim_{m \to \infty} S_{m-1}= S= \1 S.\]

So now we can simply use the familiar operation of the calculus of finite
differences. The equalities \eqref{eq:3}, \eqref{eq:4} now hold for infinite
sequences, where $\dff{1} S$ can be identify as $\oD{S}$.

\end{document}